\documentclass[%
 reprint,
 amsmath,amssymb,
 aps,
prapplied,
]{revtex4-2}

\usepackage{graphicx}
\usepackage{dcolumn}
\usepackage{bm}
\usepackage{physics}
\usepackage{amsmath,amssymb}
\usepackage{subfiles}
\usepackage{xcolor}

\newcommand{\textapprox}{\raisebox{0.5ex}{\texttildelow}}

\begin{document}


\title{Field-Based Formalism for Calculating Multi-Qubit Exchange Coupling Rates for Transmon Qubits
}

\author{Ghazi Khan}
\author{Thomas E. Roth}%
\affiliation{%
Elmore Family School of Electrical and Computer Engineering, Purdue University, Indiana 47906, USA
}%
\affiliation{Purdue Quantum Science and Engineering Institute, Purdue University, Indiana 47907, USA}




\date{\today}

\begin{abstract}
Superconducting qubits are one of the most mature platforms for achieving practical quantum computers, but significant performance improvements are still needed. To improve the engineering of these systems, 3D full-wave computational electromagnetics analyses are increasingly being turned to. Unfortunately, existing analysis approaches often rely on full-wave simulations using eigenmode solvers that are typically cumbersome, not robust, and computationally prohibitive if devices with more than a few qubits are to be analyzed. To improve the characterization of superconducting circuits while circumventing these drawbacks, this work begins the development of an alternative modeling framework that we illustrate in the context of evaluating the qubit-qubit exchange coupling rate between transmon qubits. This quantity is a key design parameter that determines the entanglement rate for fast multi-qubit gate performance and also affects decoherence sources like qubit crosstalk. Our modeling framework uses a field-based formalism in the context of macroscopic quantum electrodynamics, which we use to show that the qubit-qubit exchange coupling rate can be related to the electromagnetic dyadic Green's function linking the qubits together. We further show how the needed quantity involving the dyadic Green's function can be related to the impedance response of the system that can be easily and efficiently computed with classical computational electromagnetics tools. We demonstrate the validity and efficacy of this approach by simulating four practical multi-qubit superconducting circuits and evaluating their qubit-qubit exchange coupling rates. We validate our results against a 3D numerical diagonalization method and against experimental data where available. We also demonstrate the impact of the qubit-qubit exchange coupling rate on qubit crosstalk by simulating a multi-coupler device and identifying operating points where the qubit crosstalk becomes zero.
\end{abstract}
\maketitle
\section{Introduction}
Superconducting qubits are one of the premier candidates for realizing practical quantum computers. Among their various types, the transmon qubit \cite{koch2007charge,roth2022transmon} has largely been responsible for impressive improvements in key areas in the realm of quantum-information processing \cite{arute2019quantum,wu2021strong,jurcevic2021demonstration}, quantum sensing \cite{chen2023detecting,kristen2020amplitude}, and quantum optics \cite{puertas2019tunable,gu2017microwave}, among others. Further advancements are constantly being made, allowing for longer coherence times \cite{kjaergaard2020superconducting}, a reduction in non-ideal parasitic errors \cite{martinis2014fast,yan2018tunable,sheldon2016procedure,sundaresan2020reducing}, and improved gate fidelities gradually converging to the fault-tolerance threshold levels required for surface codes \cite{barends2014superconducting,takita2016demonstration,chen2021exponential,acharya2023suppressing}. However, to leverage the full benefits of quantum computing on practical problems, the size of quantum processors needs to rapidly scale up \cite{huang2020mermin,dalzell2020many}. This scaling presents significant engineering challenges, which is driving an increasing need to create effective modeling and optimization methods for large-scale superconducting quantum processors to address these challenges. 

One area of particular interest is the efficient modeling of design parameters governing the performance of multi-qubit gates and parasitic effects like quantum crosstalk, as these issues remain a leading error source \cite{chen2021exponential,acharya2023suppressing,proctor2022measuring}. In these systems, qubits are typically coupled together through an electromagnetic (EM) resonator operating in the dispersive regime \cite{filipp2011multimode}. In this case, the qubit-qubit exchange coupling rate is the key design parameter limiting the speed of multi-qubit gate operations \cite{krinner2020demonstration,chow2013microwave,paraoanu2006microwave}. While a large rate favors fast gate operations, it also exacerbates quantum crosstalk that limits gate performance. In particular, typical $ZZ$ quantum crosstalk scales quadratically with the exchange coupling rate for common single qubit-qubit coupler designs, which limits the fidelity of multi-qubit operations \cite{kandala2021demonstration}. Several works \cite{mundada2019suppression,kandala2021demonstration,zhao2020high} have engineered unique structures to limit $ZZ$ quantum crosstalk while trying to reduce the impact on multi-qubit gate operations. Since the qubit-qubit exchange coupling rate occurs through the multiple (formally infinite) electromagnetic (EM) modes of the device, the straightforward calculation of summing mode contributions up to some finite number to characterize these effects may be incomplete or even divergent \cite{filipp2011multimode,parra2018quantum}. As a result, there is a need to develop suitable modeling tools that can accurately and efficiently calculate key quantities like the qubit-qubit exchange coupling rate.

Thus far, several approaches have been explored to perform the general-purpose modeling of superconducting qubit devices. Early efforts focused on black-box quantization approaches \cite{nigg2012black,solgun2014blackbox}, which combine lumped element circuit quantization methods with classical circuit synthesis techniques to compute Hamiltonian characteristics using full-wave simulation tools. While this makes the approach flexible for different circuit topologies, in practice the method can demand a user-intensive curve-fitting procedure that requires performing refined simulations around the peaks of the multi-port impedance matrix of the device. Other approaches developed in \cite{minev2021energy,roth2021macroscopic,moon2024analytical} use full-wave tools to perform eigenmode decompositions of the linear part of the EM system, with the nonlinearity of the Josephson junctions introduced separately to formulate the full Hamiltonian of the system. In doing so, these methods circumvent the laborious curve-fitting processes, but can accrue a much higher computational cost due to the complexity of 3D EM eigenmode decompositions when considering the extraction of a large number of EM modes needed to model realistic systems. Further, the approaches from \cite{nigg2012black,minev2021energy} also suffer from the way the linear and nonlinear parts of the qubits in the cQED system are subdivided, which requires many Fock states to be considered per eigenmode to reach numerical convergence \cite{moon2024analytical}. All these methods are also impaired by retaining the EM modes in the Hamiltonian description, which exacerbates the exponential growth of the Hilbert space such that the size of devices that can be rigorously modeled is inherently limited.    

To alleviate many of these issues, \cite{solgun2019simple} introduced an impedance-based formalism that relates the qubit-qubit exchange coupling rate directly to the impedance response of the circuit. Recently, these methods have been extended to account for nondissipative, non-reciprocal linear coupling between qubits \cite{labarca2023toolbox}. These approaches in principle include the effects of all EM modes that can be represented by the mesh of the device used in the numerical simulation and can be performed with a significantly reduced computational cost compared to an EM eigenmode decomposition. These methods seem to work well in the qubits' straddling regime when the detuning between the qubits is smaller than their anharmonicities, but begin to diverge from ``exact'' numerical diagonalization results outside of this regime \cite{solgun2022direct}. Practically, this is a problematic limitation when modeling frequency-tunable quantum processors where it is a common strategy to strongly detune qubits from each other when not performing intentional multi-qubit gates to minimize crosstalk \cite{arute2019quantum}. It can also lead to issues when modeling certain systems designed to suppress quantum crosstalk; e.g., \cite{mundada2019suppression}. 

In this work, we use a field-based formalism in the context of macroscopic circuit quantum electrodynamics (cQED) \cite{roth2021macroscopic} to derive a formula to calculate the qubit-qubit exchange coupling rate between transmons. We show how typical field-theoretic manipulations can be used to relate the exchange coupling rate to the EM dyadic Green's function characterizing a physical device. This Green's function in principle accounts for all modes of the EM spectrum, and can be further related to the system's impedance response using standard microwave network theory \cite{pozar2021microwave,roth2022full}. Our final expressions for the exchange coupling rate are distinct from those of \cite{solgun2019simple} and do not suffer from the same limitations regarding the qubits being restricted to small detunings. We demonstrate this by simulating four realistic transmon circuits and extracting their exchange coupling rates over a wide range of frequency values and qubit detunings. We validate our results by comparing against a full-wave 3D numerical diagonalization tool \cite{moon2024analytical}.

The remainder of this work is organized in the following manner. In Section \ref{Formulation}, we present background details on the field-based formalism used in this work to establish key notations and concepts. We then manipulate this model to arrive at a simple formula expressing the qubit-qubit exchange coupling rate in terms of the EM system's impedance response. We then quantitatively and qualitatively validate the accuracy of our method in Section \ref{sec:results}, with concluding remarks made in Section \ref{sec:conclusion}. Additional derivation details related to transforming our field-based Hamiltonian to the dispersive regime are included in Appendix \ref{sec:appendix-SW} for completeness.


\section{Formulation}\label{Formulation}
\label{sec:formulation}
Before deriving our expression for the qubit-qubit exchange coupling rate, we first review key aspects of the macroscopic cQED formalism of \cite{roth2021macroscopic} in Section \ref{subsec:macroscopic-cQED}. Following this, in Section \ref{subsec:exchange-rate-formulation} we transition our system description into the dispersive regime by performing a Schrieffer-Wolff transform and specifically isolate the term related to the qubit-qubit exchange coupling rate. We then proceed to relate this term to the dyadic Green's function and later the system impedance response to yield our final result. 

\subsection{Macroscopic Circuit Quantum Electrodynamics}
\label{subsec:macroscopic-cQED}
We start from our field-based Hamiltonian approach developed in \cite{roth2021macroscopic}. This approach is performed in the framework of macroscopic QED \cite{scheel2008macroscopic} where lossless, non-dispersive media are considered in terms of macroscopic quantities like permittivity rather than through microscopic descriptions (for a simple introduction, see \cite{chew2016quantum,chew2016quantum2,chew2020quantum}). As is common in cQED, the properties of the qubits are also treated ``macroscopically'' rather than utilizing a microscopic description of the superconducting materials. Correspondingly, our starting Hamiltonian describing the general case of two transmons coupled through some EM system is given as
\begin{multline}
    \label{field-hamiltonian}
        \hat{H}=\sum_{l=1,2} \bigg( 4E_C^{(l)}(\hat{n}^{(l)})^2-E_J^{(l)}\cos\hat{\varphi}^{(l)} \bigg)  \\+\iiint\frac{1}{2}\bigg(\epsilon\hat{\textbf{E}}^2+\mu\hat{\textbf{H}}^2 \bigg) d\textbf{r}
        \\+2e\sum_{l=1,2}\iiint\hat{\textbf{E}}\cdot\textbf{d}^{(l)} \hat{n}^{(l)}d\textbf{r},
\end{multline}
where $E_C^{(l)}$ and $E_J^{(l)}$ are the charging and Josephson energies of the $l$th qubit, respectively. Further, $\hat{n}^{(l)}$ and $\hat{\varphi}^{(l)}$ are the charge and phase operators of the $l$th qubit. The second term of the Hamiltonian corresponds to the total EM energy, with $\hat{\textbf{E}}(\textbf{r},t)$ and $\hat{\textbf{H}}(\textbf{r},t)$ being the electric and magnetic field operators respectively, and $\epsilon$ and $\mu$ correspond to the permittivity and permeability of the materials throughout the system being modeled. Here, we will consider spatially-varying permittivities but assume a homogeneous permeability equal to $\mu_0$ (i.e., non-magnetic media). Finally, the last term represents the interaction between the qubits and the cavity, where $e$ is the elementary charge and $\textbf{d}^{(l)}(\textbf{r})$ parameterizes a line integration path across the Josephson junction of qubit $l$. Taking the dot product along this path with $\hat{\textbf{E}}(\textbf{r},t)$ computes the voltage seen by the Josephson junction \cite{roth2021macroscopic}. 

As shown in \cite{roth2021macroscopic}, by adopting well-defined EM approximations the field-based Hamiltonian of (\ref{field-hamiltonian}) can be consistently reduced to the more standard circuit models often found in the literature. We also note here that while we supply the expressions for the two qubit case for simplicity, the expressions can be generalized to include more qubits in a straightforward manner so that our final expressions can be used to calculate the exchange coupling between any pair of qubits in a larger circuit.

For our later derivations, it will be useful to first expand $\hat{\mathbf{E}}$ and $\hat{\mathbf{H}}$ in (\ref{field-hamiltonian}) in terms of eigenmodes of the EM wave equation. For simplicity, we will treat the quantization in a closed region terminated with a perfect electric conducting (PEC) boundary condition so that a discrete mode summation can be used, as is commonly done in quantum optics \cite{gerry2023introductory,walls2008quantum,scully1997quantum}. However, it should be noted that our final expressions can be applied in a more general setting of an open region, as is commonly the case for formulas derived in terms of EM dyadic Green's functions \cite{na2023numerical}. Now, for our particular case, we have that the EM field operators can be expanded as
\begin{align}\label{E quant}
    \hat{\textbf{E}}(\textbf{r},t)=\sum_k\sqrt{\frac{\hbar\omega_k}{2\epsilon_0}}\biggl[\hat{a}_k(t)+\hat{a}_k^\dag(t)\biggr]\textbf{E}_k(\textbf{r}),    
\end{align}
\begin{align}\label{H quant}
    \hat{\textbf{H}}(\textbf{r},t)=-i\sum_k\sqrt{\frac{\hbar\omega_k}{2\mu_0}}\biggl[\hat{a}_k(t)-\hat{a}_k^\dag(t)\biggr]\textbf{H}_k(\textbf{r}), 
\end{align}
where $\mathbf{E}_k$ and $\mathbf{H}_k$ are the electric and magnetic field eigenmodes, $\omega_k$ is the corresponding eigenfrequency, and $\hat{a}_k$ and $\hat{a}_k^\dagger$ are the annihilation and creation operators of the $k$th mode. 

We also re-express the transmon operators in terms of their energy eigenstates denoted by $\ket{j}^{(l)}$ for qubit $l$. Substituting these expressions into (\ref{field-hamiltonian}), using the orthonormality of the EM eigenmodes to simplify the spatial integrals, and applying the rotating wave approximation  eventually yields
\begin{multline}\label{original_hamiltonian}
H=\hbar\sum_{j=0}^{\infty}\sum_{l=1}^{2}q_{j}^{(l)}\ket{j}^{(l)}\bra{j}^{(l)}+\hbar\sum_{k=0}^{\infty}\omega_k\hat{a}_k^\dag \hat{a}_k\\
+\sum_{j,k=0}^{\infty}\sum_{l=1}^2\hbar\biggl[g_{j,k}^{(l)}\ket{j}^{(l)}\bra{j+1}^{(l)}\hat{a}_k^\dag+\textnormal{H.c.}\biggr].\\
\end{multline}
In (\ref{original_hamiltonian}), $q_{j}^{(l)}$ denotes the eigenfrequency of the $j$th level of qubit $l$. We have further simplified the interaction term by assuming that only the nearest-neighbor transitions between $j\leftrightarrow j+1$ levels that are dominant in transmon qubits \cite{koch2007charge} need to be considered. Finally, we can explicitly find that within this model the coupling between a specific qubit and EM mode is given by
\begin{multline}
\label{g's expression}   
g_{j,k}^{(l)}=2e\bra{j}^{(l)}\hat{n}^{(l)}\ket{j+1}^{(l)}\cross\\\sqrt{\frac{\omega_k}{2\epsilon_0\hbar}}\iiint {\textbf{E}}_k({\textbf{r}})\cdot {\textbf{d}}^{(l)}({\textbf{r}})d{\textbf{r}}.
\end{multline}

\subsection{Qubit-Qubit Exchange Coupling Rate}
\label{subsec:exchange-rate-formulation}
Superconducting qubits are usually operated in the dispersive regime, which is characterized by having the detuning between the EM resonant modes and qubits far stronger than their corresponding coupling strength; i.e., $g_{j,k}^{(l)}/\Delta^{(l)}_{j,k}\ll 1$, where $\Delta^{(l)}_{jk}=q_{j,j+1}^{(l)}-\omega_k$ and $q_{j,j+1}^{(l)}=q_{j+1}^{(l)}-q_{j}^{(l)}$. In this regime, the coupled EM-qubit system interacts through virtual processes, which has many benefits for qubit readout and control \cite{krantz2019quantum,blais2021circuit}. However, since virtual processes often involve higher energy levels, it is essential to account for the full multi-level transmon structure as we have done in (\ref{original_hamiltonian}) \cite{blais2021circuit}. 

To more effectively study the system in the dispersive regime, it is common to perform a Schrieffer-Wolff transform \cite{zhu2013circuit} to express the system Hamiltonian in a form that makes the key dispersive regime physics explicit. The exact details of the transform have been included in Appendix \ref{sec:appendix-SW}. The key result is that this transform rewrites the indirect coupling of qubits through their mutual interactions with EM fields into direct qubit-qubit interactions with EM-dependent coupling strengths. This new effective interaction term corresponds to the last term of (\ref{sw-og}), which is reproduced here and denoted as $V'$ for clarity. The term is
\begin{multline}
    V'=\sum_{i,j} J_{ij}(\ket{i}^{(1)}\bra{i+1}^{(1)}\ket{j+1}^{(2)}\bra{j}^{(2)}\\+\ket{i+1}^{(1)}\bra{i}^{(1)}\ket{j}^{(2)}\bra{j+1}^{(2)}),
\end{multline}
where
\begin{align}
\label{SW-Interaction Hamiltonian}
    J_{ij}= \sum_k \frac{\hbar}{2} \left[\frac{{g}_{i,k}^{*(1)}{g}_{j,k}^{(2)}}{(q_{i,i+1}^{(1)}-\omega_k)}+\frac{{g}_{i,k}^{(1)}{g}_{j,k}^{*(2)}}{(q_{j,j+1}^{(2)}-\omega_k)}\right]
\end{align}
is the qubit-qubit exchange coupling rate.

To, improve the clarity of the derivation we will split (\ref{SW-Interaction Hamiltonian}) into two terms defined as
\begin{align}\label{First term of J}
    J_{ij}^{\{1\}}= \sum_k\frac{\hbar}{2} \frac{{g}_{i,k}^{*(1)}{g}_{j,k}^{(2)}}{(q_{i,i+1}^{(1)}-\omega_k)},
\end{align}
\begin{align}\label{Second term of J}
    J_{ij}^{\{2\}}= \sum_k \frac{\hbar}{2} \frac{{g}_{i,k}^{(1)}{g}_{j,k}^{*(2)}}{(q_{j,j+1}^{(2)}-\omega_k)},
\end{align}
and focus on the first; i.e., (\ref{First term of J}). We expand this term using the definitions from (\ref{g's expression}) to get
\begin{multline}
    J_{ij}^{\{1\}}=\sum_k\biggl[\frac{e^2}{\epsilon_0}\bra{i+1}^{(1)}|\hat{n}^{(1)}\ket{i}^{(1)}\bra{j}^{(2)}|\hat{n}^{(2)}\ket{j+1}^{(2)}\\
    \cross\frac{\omega_k\iint\textbf{d}^{(1)}(\textbf{r})\cdot \textbf{E}_k(\textbf{r})\textbf{E}_k(\textbf{r}')\cdot\textbf{d}^{(2)}(\textbf{r}')d\textbf{r}d\textbf{r}'}{q^{(1)}_{i,i+1}-\omega_k}\biggr]
\end{multline}
We can approximate the summation over EM modes with an integration over a continuous eigenfrequency $\omega$ to get
\begin{multline}\label{J expand}
    J_{ij}^{\{1\}}=\frac{e^2}{\epsilon_0}\hat{n}^{(1)}_{i+1,i}\hat{n}^{(2)}_{j,j+1}\cross\\\int
    \frac{\omega\rho (\omega)\iint\textbf{d}^{(1)}(\textbf{r})\cdot \textbf{E}(\textbf{r},\omega)\textbf{E}(\textbf{r}',\omega)\cdot\textbf{d}^{(2)}(\textbf{r}')d\textbf{r}d\textbf{r}'}{q^{(1)}_{i,i+1}-\omega}d\omega
\end{multline}
where $\rho(\omega)$ is the EM density of states and we have introduced the shorthand $n_{i,j}^{(l)} = \bra{i}^{(l)}\hat{n}^{(l)}\ket{j}^{(l)}$ to simplify the notation. A common result from macroscopic QED is to relate the local EM density of states to the EM dyadic Green's function as \cite{liu2017dressed}
\begin{align}\label{density of states}
    \rho(\omega)\frac{\omega}{2\epsilon_0}\textbf{E}(\textbf{r},\omega)\textbf{E}(\textbf{r}',\omega)=\frac{\omega^2\mu_0}{\pi}\Im{ \bar{\textbf{G}}(\textbf{r},\textbf{r}',\omega)}.
\end{align}
Using (\ref{density of states}) in (\ref{J expand}), we get
\begin{multline}\label{J with green}
        J_{ij}^{\textnormal{\{1\}}}=\frac{2}{\pi}e^2\mu\hat{n}_{i+1,i}^{(1)}\hat{n}_{j,j+1}^{(2)}\cross\\
         \int\frac{\omega^2\iint{\textbf{d}}^{(1)}({\textbf{r}})\cdot \mathrm{Im}\{ \bar{\textbf{G}}(\textbf{r},\textbf{r}',\omega) \} \cdot\textbf{d}^{(2)}(\textbf{r}{'})d\textbf{r}d\textbf{r}{'}}{q_{i,i+1}^{(1)}-\omega}d\omega.
\end{multline}

With full details given in \cite{roth2022full}, we now evaluate the integral over the Green's function to relate it to the system impedance response using results from microwave network theory. In particular, we can consider the effective two-port network formed between the qubits in the system. If we drive qubit 2 with an impressed current given by $-\textbf{d}^{(2)}(\textbf{r}')I^{(2)}_t$, we can determine that the open circuit voltage induced across qubit 1 will be
\begin{multline} \label{Intermediate to Z}
V_t^{(1)} = \int\textbf{d}^{(1)}(\textbf{r})\cdot\textbf{E}(\textbf{r})d\textbf{r} =    -i\omega\mu_0 I^{(2)}_t  \\ \times \iint\textbf{d}^{(1)}(\textbf{r})\cdot\bar{\textbf{G}}(\textbf{r},\textbf{r}',\omega)\cdot\textbf{d}^{(2)}(\textbf{r}')d\textbf{r}d\textbf{r}' .
\end{multline}
Then, we can rearrange this into the definition of a transfer impedance as $Z_{12}=V^{(1)}_t/I^{(2)}_t$ \cite{pozar2021microwave}. Finally, we can take the imaginary part of both sides of (\ref{Intermediate to Z}) to get a succinct expression of
\begin{align}\label{green to impedence}
    \iint\textbf{d}^{(1)}(\textbf{r})\cdot \mathrm{Im}\{\bar{\textbf{G}}(\textbf{r},\textbf{r}{'},\omega)\}\cdot&\textbf{d}^{(2)}(\textbf{r}{'})d\textbf{r}d\textbf{r}'\nonumber\\&=\frac{\Re{Z_{12}(\omega)}}{\omega\mu_0},
\end{align}
where we have further noted that $\Im{iZ_{12}}=\Re{Z_{12}}$. It is important to note that for now we will assume an infinitesimal amount of loss in the system, but will take the limit to zero in the end to match our initial lossless assumptions about the system. 

We can now use (\ref{green to impedence}) in (\ref{J with green}) to replace the Green's function with the impedance response of the system to get
\begin{align}\label{J with impedance}
        J_{ij}^{\textnormal{\{1\}}}=\frac{2}{\pi}e^2\hat{n}_{i+1,i}^{(1)}\hat{n}_{j,j+1}^{(2)}
         \int\frac{\omega\Re{Z_{12}(\omega)}}{q_{i,i+1}^{(1)}-\omega}d\omega.
\end{align}
The integration over $\omega$ can be handled using the Plemelj formula \cite{stone2009mathematics}, which expresses the integration as  
\begin{multline}
        J_{ij}^{\textnormal{\{1\}}}={2}e^2\hat{n}_{i+1,i}^{(1)}\hat{n}_{j,j+1}^{(2)}
         \cross\\\biggl[{\frac{1}{\pi}{\textrm{P.V.}}\biggl[\frac{\omega\Re{Z_{12}(\omega)}}{q_{i,i+1}^{(1)}-\omega}\biggr]}-iq_{i,i+1}^{(1)}\Re{Z_{12}(q_{i,i+1}^{(1)})}\biggr].
         \label{eq:intermed-J}
\end{multline}
The remaining principal value integration can be readily seen to be a Hilbert transform, which can be evaluated easily due to the analyticity of $\omega$ and $\Re{Z_{12}(\omega)}$. In particular, if we denote the Hilbert transform as
\begin{align}
    \mathcal{H}\bigl\{\omega\Re{Z_{12}(\omega)}\bigr\}(q_{i,i+1}^{(1)}) = {\frac{1}{\pi}{\textrm{P.V.}}\biggl[\frac{\omega\Re{Z_{12}(\omega)}}{q_{i,i+1}^{(1)}-\omega}\biggr]},
\end{align}
we can use a relevant product theorem of the Hilbert transform of analytic functions \cite{bedrosian1962product} to simplify to
\begin{multline}
    \mathcal{H}\bigl\{\omega\Re{Z_{12}(\omega)}\bigr\} (q_{i,i+1}^{(1)}) \\ = q_{i,i+1}^{(1)}\mathcal{H}\bigl\{\Re{Z_{12}(\omega)}\bigr\}(q_{i,i+1}^{(1)}).
\end{multline}
Finally, the Hilbert transform can be evaluated to get
\begin{align}
    \mathcal{H}\bigl\{\omega\Re{Z_{12}(\omega)}\bigr\} (q_{i,i+1}^{(1)})  = q_{i,i+1}^{(1)}\Im{Z_{12}(q^{(1)}_{i,i+1})}.
\end{align}

Using this result in (\ref{eq:intermed-J}), we get the complete expression for $J_{ij}^{\{1\}}$ to be
\begin{multline}
        J_{ij}^{\textnormal{\{1\}}}={2}e^2\hat{n}_{i+1,i}^{(1)}\hat{n}_{j,j+1}^{(2)}
         \cross\\\biggl[q_{i,i+1}^{(1)}\Im{Z_{12}(q_{i,i+1}^{(1)})}\!-\!iq_{i,i+1}^{(1)}\Re{Z_{12}(q_{i,i+1}^{(1)})}\biggr].
\end{multline}
At this point, we take the limit of ``vanishing loss'' in the EM system so that $\Re{Z_{12}}=0$. Performing a similar set of steps for the second term of $J_{ij}$ given in (\ref{Second term of J}), we get the complete qubit-qubit exchange coupling rate to be 
\begin{multline}
            J_{ij}=2e^2
        \biggl[n_{i+1,i}^{(1)}n_{j,j+1}^{(2)} q_{i,i+1}^{(1)}\Im{ Z_{12}(q_{i,i+1}^{(1)})}
        \\+n_{j+1,j}^{(2)} n_{i,i+1}^{(1)}q_{j,j+1}^{(2)}\Im{Z_{21}(q_{j,j+1}^{(2)})}\biggr].
\end{multline}

Typically, the specific qubit-qubit exchange coupling rate of interest is for the coupling between the first 2 levels of the transmons. Therefore we will evaluate $J_{ij}$ at $i,j=0$ and redefine the quantity as $J$ to better match the common notation used in the literature. The full term that will be evaluated in our validation in Section \ref{sec:results} is then 
\begin{multline}\label{Final J}
    J=2e^2 \biggl[{n}^{(1)}_{1,0}{n}^{(2)}_{0,1} q_{0,1}^{(1)}\textnormal{Im}\{ Z_{12}(q_{0,1}^{(1)}) \}\\+{n}^{(2)}_{1,0}{n}^{(1)}_{0,1}q_{0,1}^{(2)}\textnormal{Im}\{Z_{21}(q_{0,1}^{(2)})\}\biggr].
\end{multline} 

\section{Numerical Results}\label{sec:results}
In this section, we validate our approach for calculating the qubit-qubit exchange coupling rate through simulating various circuit topologies. We use Ansys HFSS to perform full-wave finite element simulations to compute the needed impedance parameters. As alluded to in our formulation, we perform these impedance simulations by replacing the Josephson junctions in each topology with lumped ports that the impedance parameters can be computed between. It should also be noted that although our initial derivation was performed under the assumption of a large enclosed region (justifying the continuum approximation made in (\ref{J expand})), performing an explicit simulation of such a setup is computationally prohibitive and unnecessary. Here, we use a radiation boundary condition on all outer surfaces of our simulation region placed an appropriate distance from our device. We further note that the planar devices are modeled as infinitely thin PEC sheets. While in practice these conductors have a finite thickness of $O(100 \, \mathrm{nm})$, explicitly modeling such thin conductors substantially increases the computational time and is unnecessary to demonstrate the validity of our overall formulation. Finally, to compute the charge matrix elements and qubit frequencies we use the simple numerical method described in \cite{roth2024maxwell}. The qubit capacitances are computed from the input impedance of each qubit lumped port in our HFSS simulations.

This remainder of this section is organized in the following way. In Section \ref{subsec:rwg}, we analyze the qubit-qubit exchange coupling rate for two transmons embedded in a 3D waveguide cavity. We repeat a similar analysis in Section \ref{subsec:cpw-wallraff} for a planar design qualitatively adapted from \cite{filipp2011multimode}. To demonstrate the flexibility of our approach, we next model a four qubit planar device with direct capacitive coupling from \cite{du2024probing} in Section \ref{subsec:4-qubit_Ma}. Finally, in Section \ref{subsec:Houck} we model a planar device based on \cite{mundada2019suppression} that uses a multi-coupler architecture to cancel the $ZZ$ crosstalk between two qubits without nulling their desirable qubit-qubit exchange coupling rate.


\subsection{3D Rectangular Waveguide Cavity}
\label{subsec:rwg}
\begin{figure}[t]
    \centering
    \includegraphics[width=\linewidth]{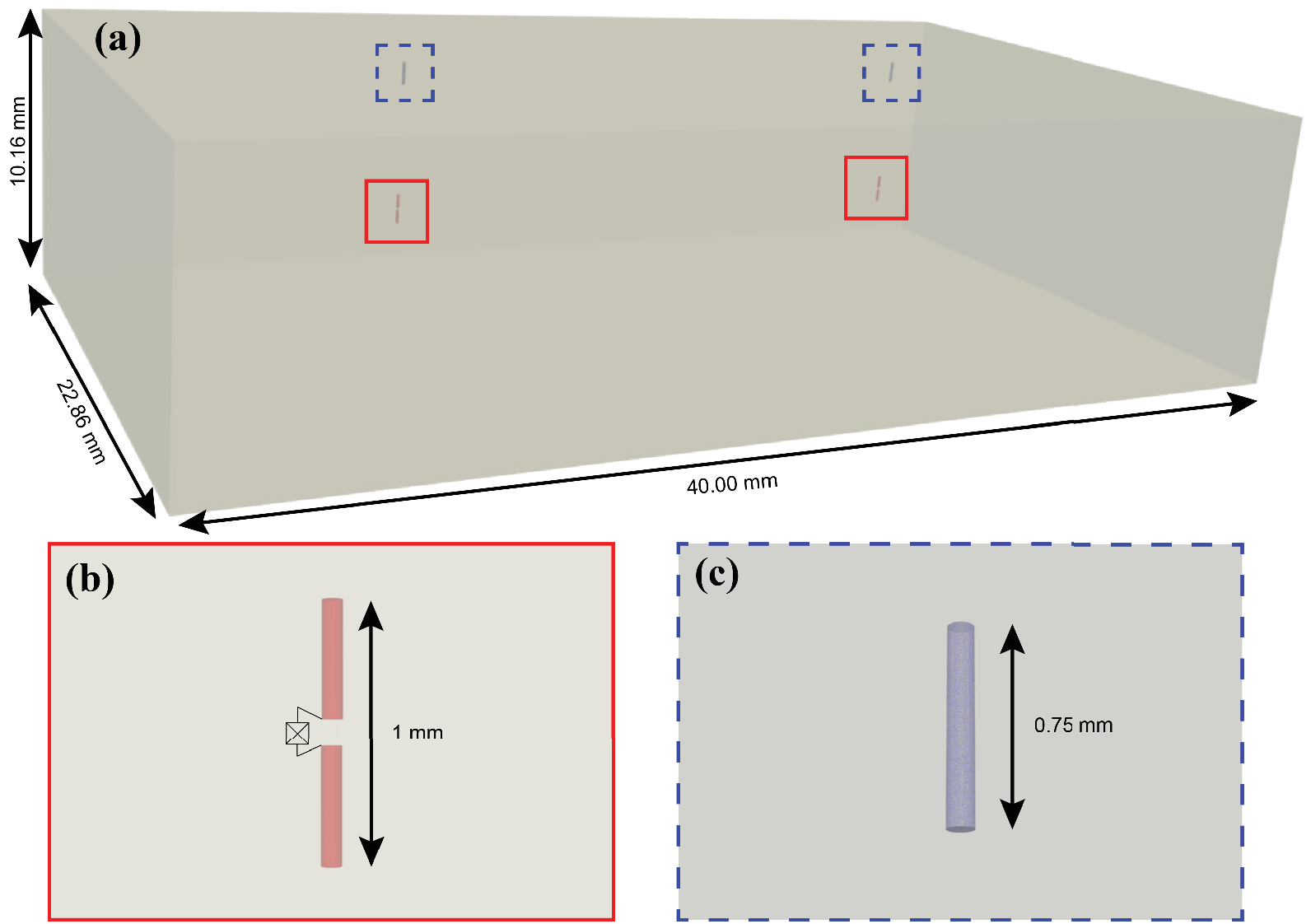}
    \caption{(color online) Section \ref{subsec:rwg} Device: (a) Image of 3D waveguide cavity with two transmons (solid red boxes) and two tuning probes (dashed blue boxes). (b) Closeup view of a transmon formed by a short wire dipole antenna with a Josephson junction connected across its terminals.  (c) Closeup view of one of the short tuning probes.}
    \label{Soomin_geometry}
\end{figure}

The first system we analyze is a 3D rectangular waveguide cavity that contains two tuning probes and two transmons formed by connecting Josephson junctions across the terminals of small wire dipole antennas, as shown in Fig. \ref{Soomin_geometry}. This system geometry was designed specifically so that a full-wave analytical solution could be formulated \cite{moon2024analytical}, but here we only focus on numerical solutions. The dimensions have been set so that the fundamental frequency of the cavity is 7.55 $\mathrm{GHz}$. The dipole terminals are also loaded with a $50 \, \mathrm{fF}$ capacitance to place their operating point in the transmon regime. Our simulations then give a final qubit capacitance $C_1,\,C_2=57.24\,\mathrm{fF}$
. Finally, the entire cavity region outside of the PEC objects have permittivity and permeability set to their respective free space values.

\begin{figure}[t]
    \centering
    \includegraphics[width=0.9\linewidth]{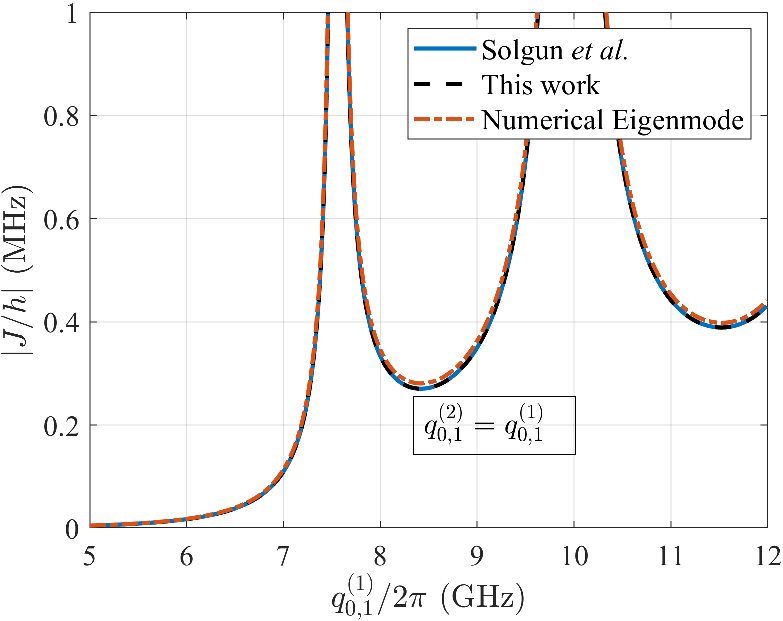}
    \caption{Section \ref{subsec:rwg} Device: Calculation of $J$ for a sweep across qubit frequencies, while keeping both equal.}
    \label{Soomin_geometry_equal}
\end{figure}

We evaluate the qubit-qubit exchange coupling rate using (\ref{Final J}). First, we perform a sweep for a wide range of frequencies while keeping both qubits equal in frequency. The results are shown in Fig. \ref{Soomin_geometry_equal}, where we see excellent agreement when compared to the impedance-based method of Solgun \textit{et al.} \cite{solgun2019simple} and the 3D numerical diagonalization method from \cite{moon2024analytical}.  

\begin{figure}[t]
    \centering
    \includegraphics[width=0.9\linewidth]{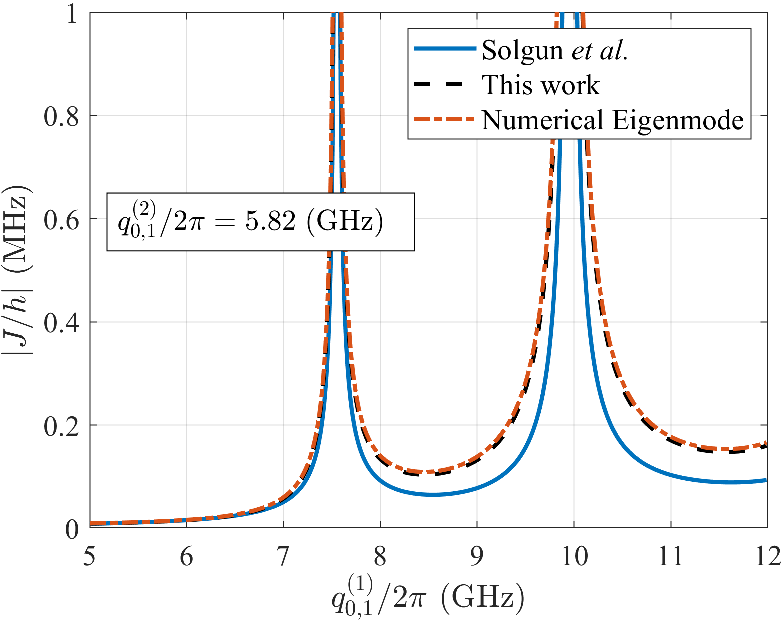}
    \caption{Section \ref{subsec:rwg} Device: Calculation of $J$ while keeping one qubit constant at 5.82 $\mathrm{GHz}$ while sweeping across the other qubit frequency.}
    \label{Soomin_geometry_5_82}
\end{figure}

\begin{figure}[t]
    \centering
    \includegraphics[width=0.9\linewidth]{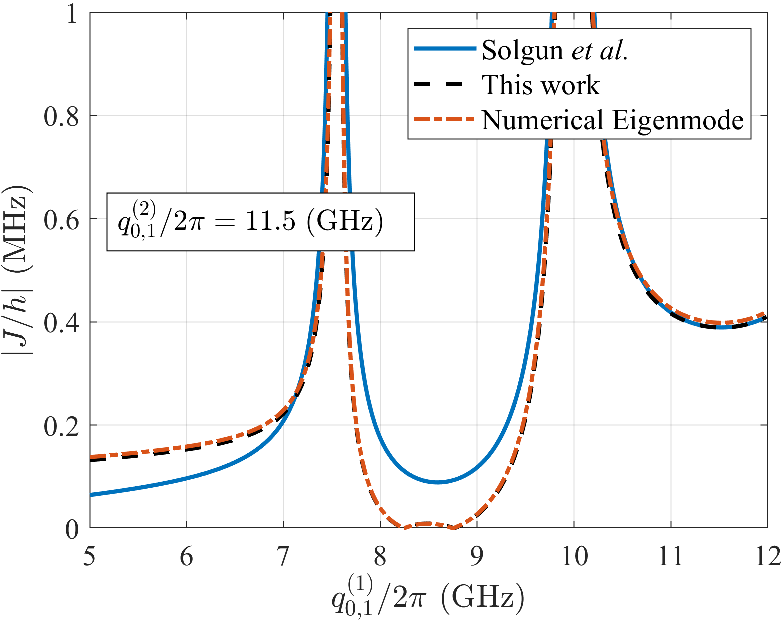}
    \caption{Section \ref{subsec:rwg} Device: Calculation of $J$ while keeping one qubit constant at 11.5 $\mathrm{GHz}$ while sweeping across the other qubit frequency.}
    \label{Soomin_geometry_11_5}
\end{figure}

We then repeat the process but this time fix the frequency of one qubit and sweep across the frequencies of the other qubit. We do this once for the fixed qubit frequency at 5.82 $\mathrm{GHz}$ so that it is below the fundamental cavity mode before then changing the fixed qubit frequency to 11.5 $\mathrm{GHz}$ so that it is above the second cavity mode. The results for these two cases are shown  in Figs. \ref{Soomin_geometry_5_82} and \ref{Soomin_geometry_11_5}, respectively. In both cases, we see that while the two qubits are close in frequency, all three methods agree. However, when the two qubits are detuned from each other by a few $\mathrm{GHz}$ the method from \cite{solgun2019simple} no longer agrees with the 3D numerical diagonalization while our method continues to be highly accurate. This accuracy over a wider range of qubit detunings is important to enhance the robustness of the numerical predictions, and is particularly important for flux-tunable processors where strongly detuning qubits is a common strategy to mitigate crosstalk during idle periods \cite{arute2019quantum}.

\subsection{Coplanar Waveguide Resonator}
\label{subsec:cpw-wallraff}
\begin{figure}[t]
    \centering
    \includegraphics[width=\linewidth]{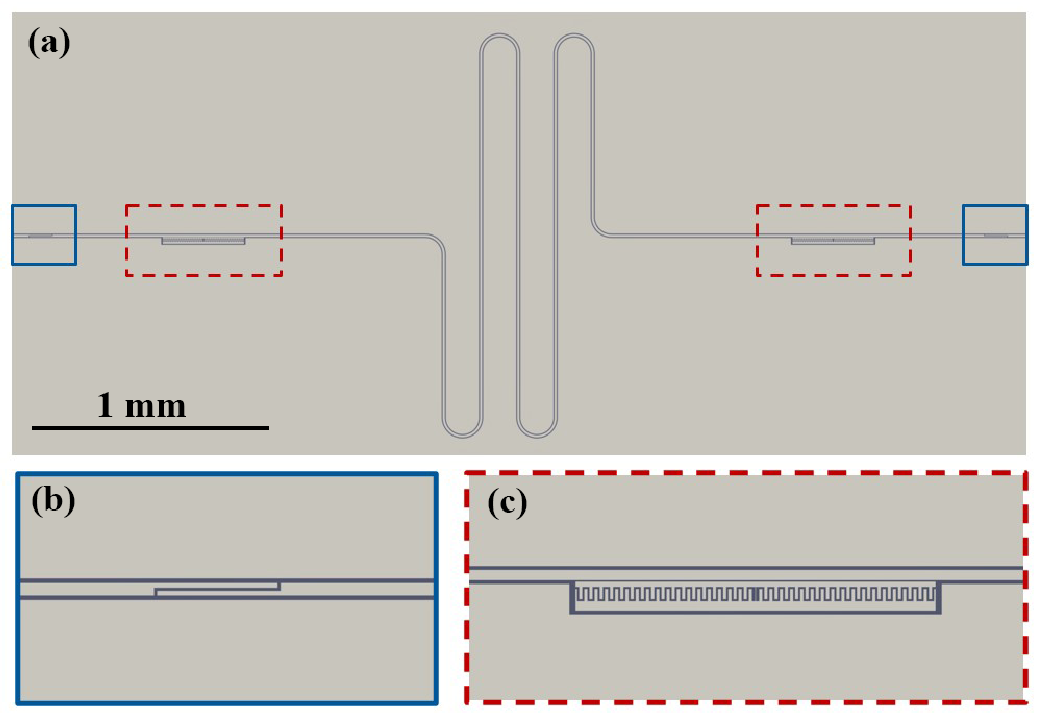}
    \caption{(color online) Section \ref{subsec:cpw-wallraff} Device: (a) Half-wavelength coplanar waveguide resonator formed between two interdigital capacitors (solid blue boxes) coupled capacitively to two transmons (red-dotted boxes). (b) Closeup of interdigital capacitive coupling at the end of the coplanar waveguide resonator. (c) Closeup of a transmon formed by a large interdigital capacitor.}
    \label{Walraff_design}
\end{figure}

The next design considered is based on the device studied experimentally in \cite{filipp2011multimode}, with our corresponding model shown in Fig. \ref{Walraff_design}. This device consists of two transmons capacitively coupled to a half-wavelength coplanar waveguide resonator. The coplanar waveguide resonator is sized so that the fundamental mode occurs at 6.33 $\mathrm{GHz}$. The qubit capacitances are simulated to be $C_1,=69.62\,\mathrm{fF}$ and $\,C_2=69.63\,\mathrm{fF}$. Instead of simulating a finite thickness substrate, we simplify our simulation model here by using an effective relative permittivity of $\epsilon_\mathrm{eff}=4.945$ throughout the entire simulation domain. This effective permittivity is computed assuming a thick silicon substrate using typical formulas \cite{simons2004coplanar}. 

\begin{figure}[t]
    \centering
    \includegraphics[width=0.9\linewidth]{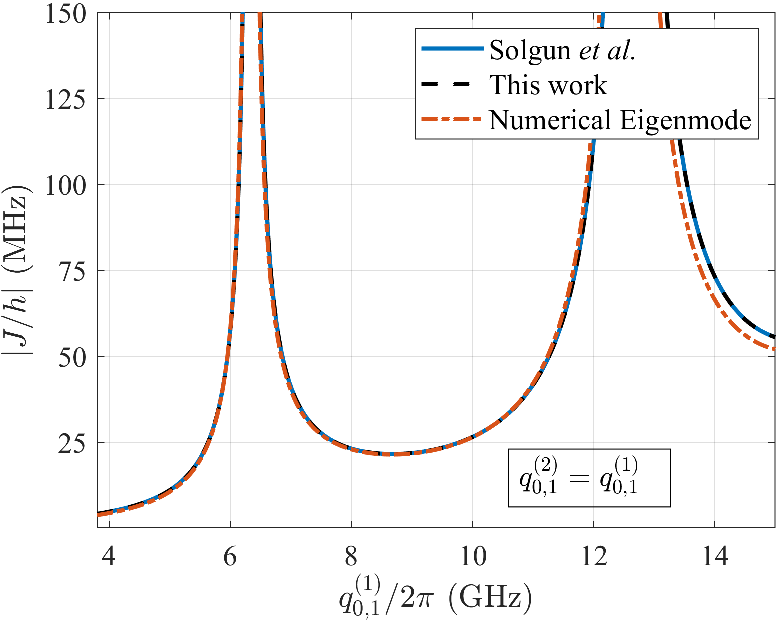}
    \caption{Section \ref{subsec:cpw-wallraff} Device: Calculation of $J$ for a sweep across qubit frequencies, while keeping both equal.}
    \label{Walraff_equal}
\end{figure}

\begin{figure}[t]
    \centering
    \includegraphics[width=0.9\linewidth]{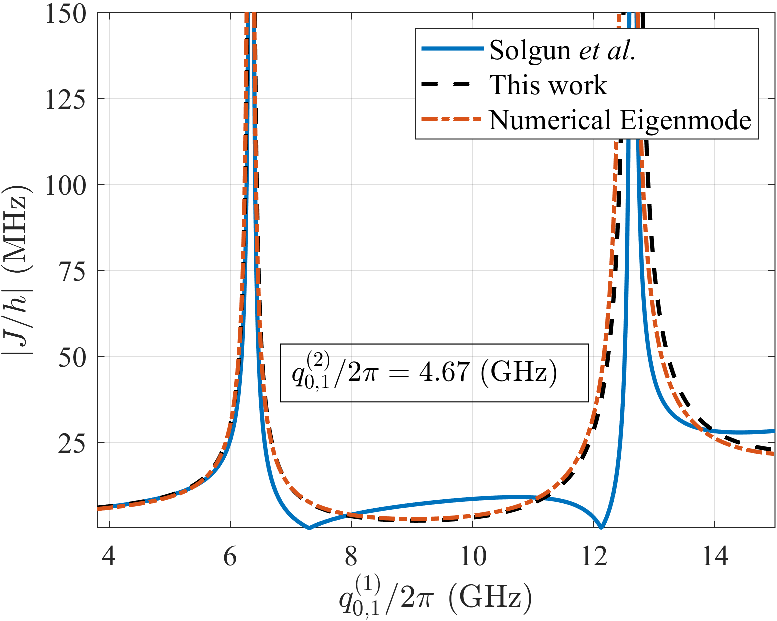}
    \caption{Section \ref{subsec:cpw-wallraff} Device: Calculation of J while keeping one qubit constant at 4.67 GHz while sweeping across the other qubit frequency.}
    \label{Walraff_4_67}
\end{figure}

\begin{figure}[t]
    \centering
    \includegraphics[width=0.9\linewidth]{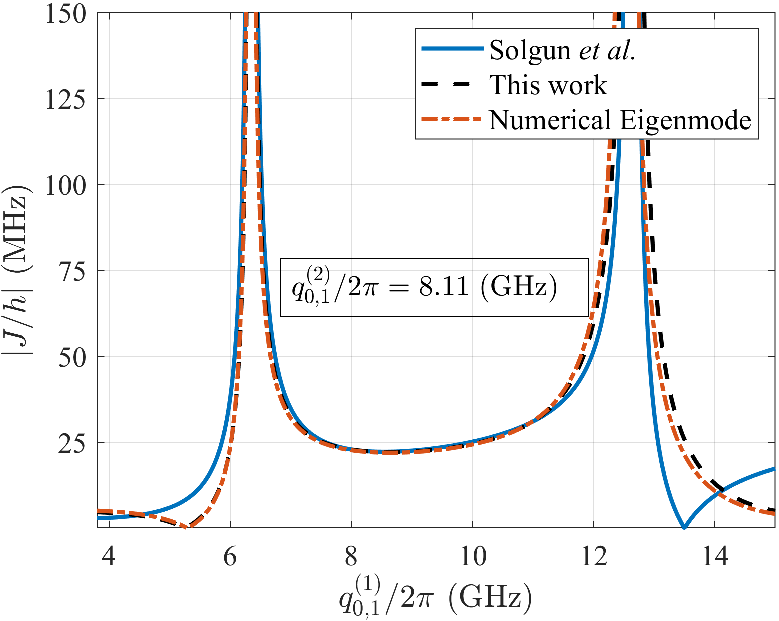}
    \caption{Section \ref{subsec:cpw-wallraff} Device: Calculation of J while keeping one qubit constant at 8.11 GHz while sweeping across the other qubit frequency.}
    \label{Walraff_8_11}
\end{figure}

Here, we also have computed the qubit-qubit exchange coupling rate for the case of equal qubit frequencies (Fig. \ref{Walraff_equal}), one qubit fixed below the fundamental mode at 4.67 $\mathrm{GHz}$ (Fig. \ref{Walraff_4_67}), and one above the fundamental mode at 8.11 $\mathrm{GHz}$ (Fig. \ref{Walraff_8_11}). We see similar behavior as to the results in Section \ref{subsec:rwg}. That is, when the qubits are equal in frequency, the impedance-based formalism of our work and Solgun \textit{et al.} \cite{solgun2019simple} agree with the 3D numerical diagonalization method of \cite{moon2024analytical}; however, when the qubits are sufficiently detuned from each other, the method from \cite{solgun2019simple} becomes inaccurate while our method suffers no such issues. We also see good qualitative agreement with our results and the experimental results from \cite{filipp2011multimode}.

\subsection{Planar Device with Direct Capacitive Coupling}
\label{subsec:4-qubit_Ma}
\begin{figure}[t]
    \centering
    \includegraphics[width=\linewidth]{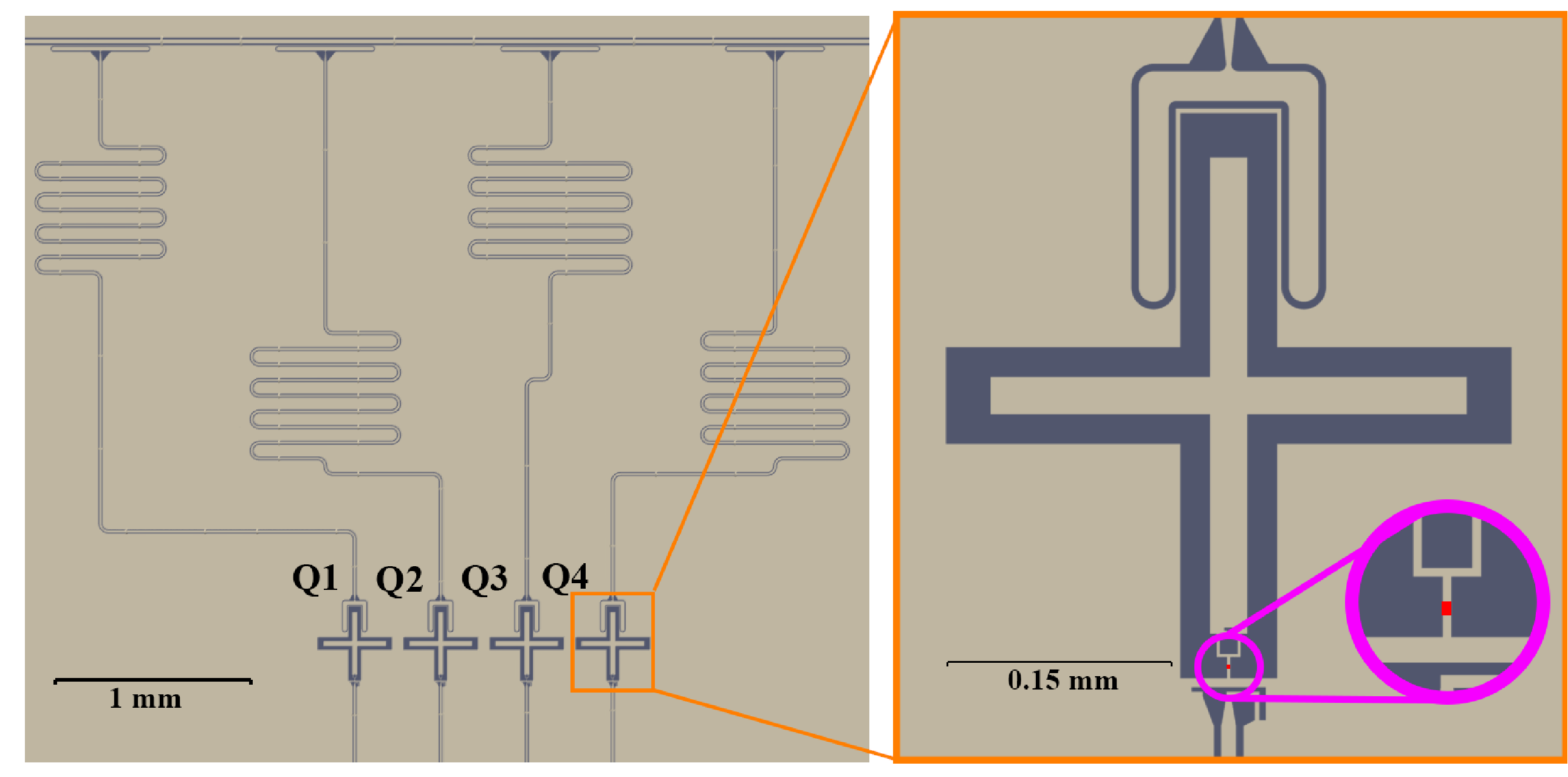}
    \caption{(color online) Section \ref{subsec:4-qubit_Ma} Device: Four transmon device with capacitive coupling to nearest-neighbors and to independent readout resonators. The magnification on the SQUID loop (magenta circle) shows how it is excited using a single lumped port, denoted by a red box.}
    \label{Ma_design}
\end{figure}

Here, we model the device from \cite{du2024probing} that has four transmon qubits with nearest-neighbor direct capacitive coupling, as shown in Fig. \ref{Ma_design}. Each transmon also has its own half-wavelength readout resonator, which have fundamental resonant frequencies in the range of $6.2-6.4\,\mathrm{GHz}$. Here, we model the device on top of a $0.5 \, \mathrm{mm}$ thick sapphire substrate. To simplify our model, we excite the superconducting quantum interference device (SQUID) loop shown in Fig. \ref{Ma_design} with a single lumped port. Due to the highly subwavelength nature of the SQUID loop, we do not expect this simplification to significantly impact the impedance parameters. Finally, in contrast to the previous devices, here our HFSS simulation model is built from the design files explicitly used in building the device of \cite{du2024probing} (with the exception of the airbridges added into the model) which enables direct quantitative comparison to their experimental results.  

\begin{figure}[t]
    \centering
    \includegraphics[width=0.9\linewidth]{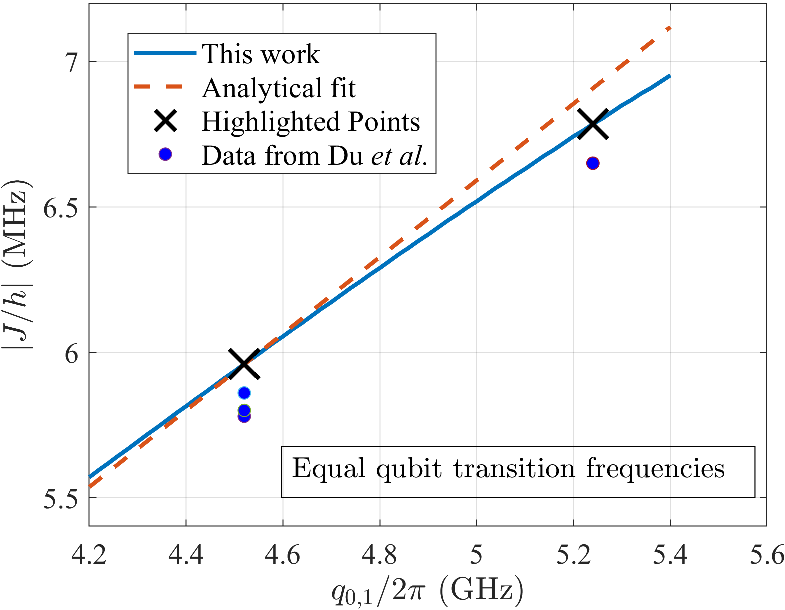}
    \caption{Section \ref{subsec:4-qubit_Ma} Device: Calculation of $J$ using (\ref{Final J}) for a sweep across quibt frequencies, while keeping all frequencies equal. Also shown is the analytical fit using (\ref{eq:analytical-J}). We highlight points with an ``$\cross$'' for which we have experimental data from \cite{du2024probing}. The measured results from \cite{du2024probing} are also shown, exhibiting a relative error ranging from \textapprox1 to 4\%.}
    \label{Ma_design_Results}
\end{figure}

As with previous devices, we perform our impedance simulations to evaluate (\ref{Final J}) and present the results in Fig. \ref{Ma_design_Results}. It should be noted that the direct capacitive coupling of qubits in this device prevents the use of the 3D numerical diagonalization method of \cite{moon2024analytical} without further modifying it; however, our impedance-based formalism is able to be applied with no modifications. From Fig. \ref{Ma_design_Results}, we see good agreement between our approach and the measured values, with relative errors ranging in the \textapprox1 to 4\% range. We attribute this difference primarily to numerical error, which we expect is dominated by our modeling the actual $150\,\mathrm{nm}$ Niobium layer as an infinitely thin PEC sheet. We also only report the $J$ between qubits 1 and 2, since our numerical calculations showed a very small difference of \textapprox$0.004\,\mathrm{MHz}$ between this $J$ and that of the other nearest-neighbor qubit pairs. 

To better place the observed numerical errors in perspective, we also use the typical analytical formula for calculating the qubit-qubit exchange coupling rate due to direct capacitive coupling between qubits to analyze this system. In particular, the analytical formula is \cite{yan2018tunable} 
\begin{align}
    J/\hbar\approx\frac{1}{2}\frac{C_c}{\sqrt{C_1C_2}}\sqrt{q^{(1)}_{0,1}  q^{(2)}_{0,1}},
    \label{eq:analytical-J}
\end{align}
where $C_c$ is the coupling capacitance between qubits 1 and 2 and $C_1$ and $C_2$ are the total qubit capacitances to ground. We then determine the value of $C_c$ needed to fit our numerical data at the main 4.52 GHz operating frequency of the device from \cite{du2024probing} when using the values of $C_1=81.94\,\mathrm{fF}$ and $C_2=81.93\,\mathrm{fF}$ determined from our impedance simulations, which results in $C_c = 0.216 \, \mathrm{fF}$ (results of this fit are also shown in Fig. \ref{Ma_design_Results}). We then repeat this exercise to determine that a value of $C_c = 0.209 \, \mathrm{fF}$ would fit the experimental data well. This shift of $C_c$ by $0.07\,\mathrm{fF}$ is very small compared to the other capacitances present in the system, and is reasonable to be at least partially attributed to numerical error (past studies on finite thickness effects, e.g., \cite{tran1993full}, suggest this would lower $C_c$). However, given the small scale of the capacitances involved, manufacturing tolerances and other experimental effects could also account for some of the observed deviation between numerical and experimental results.

\subsection{Planar Multi-Coupler Device for $ZZ$ Cancellation}
\label{subsec:Houck}
\begin{figure}[t]
    \centering
    \includegraphics[width=\linewidth]{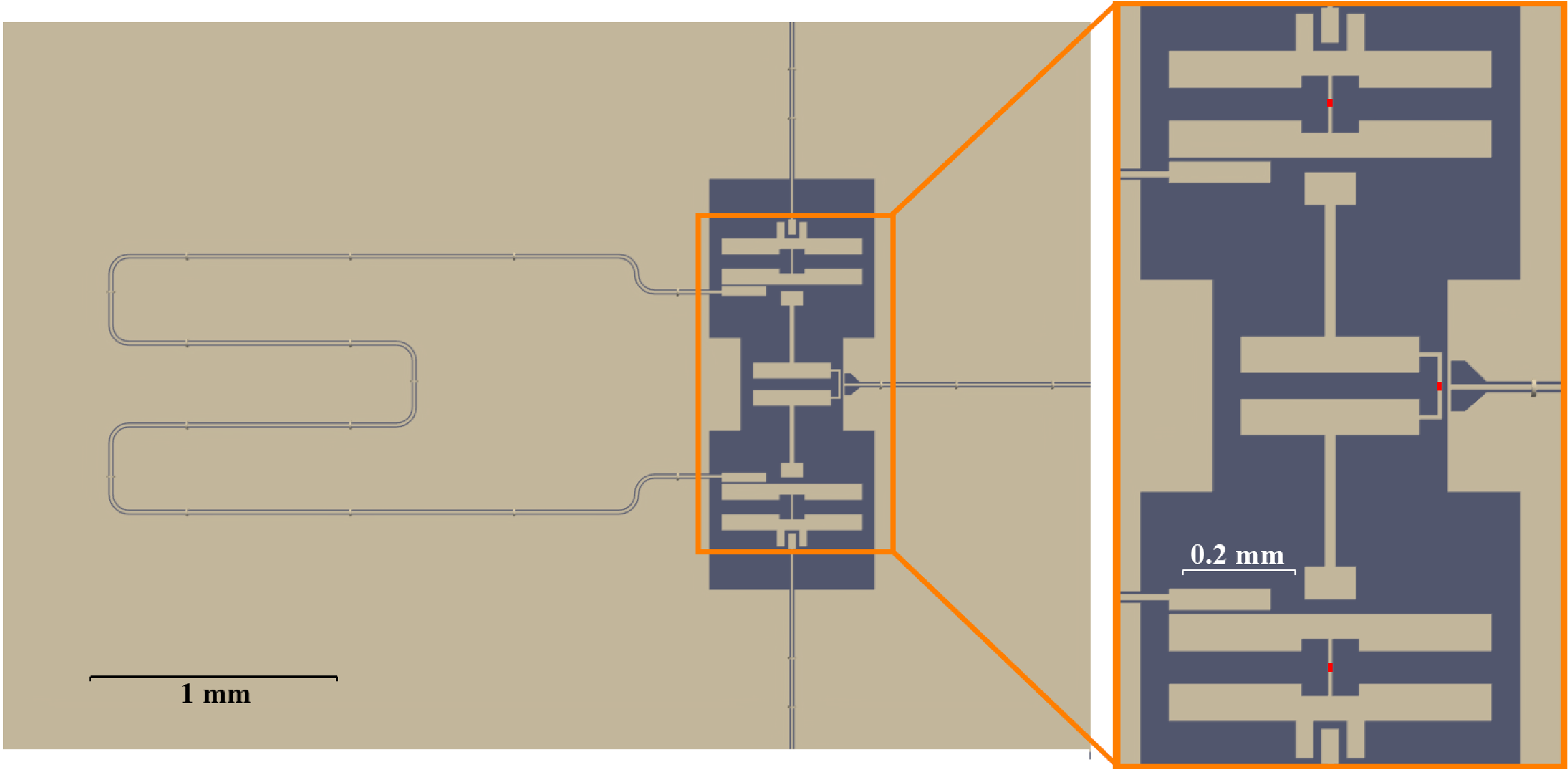}
    \caption{(color online) Section \ref{subsec:Houck} Device: Planar device that utilizes a dual-coupling approach where the qubits are coupled through the distributed resonator and the tunable SQUID coupler to accomplish $ZZ$ crosstalk cancellation. In the context of our simulations, both qubits and SQUID coupler are modeled as single lumped elements denoted by the red boxes.}
    \label{Houck_design}
\end{figure}

The final design considered is based on the device studied experimentally in \cite{mundada2019suppression}, with our corresponding model shown in Fig. \ref{Houck_design}. This device consists of two transmon qubits, which are coupled to each other through two different resonators. One is a fixed half-wavelength coplanar waveguide resonator and the other is an additional transmon with a SQUID to facilitate tuning (referred to as the SQUID coupler). As with the previous device, we simplify our model by replacing the SQUID loop with a single lumped port, as shown in Fig. \ref{Houck_design}. This multi-coupler approach is designed so that a usable $J$ rate can be achieved to facilitate multi-qubit operations while simultaneously reducing the $ZZ$ crosstalk that is a problematic source of decoherence in such devices.

To numerically model this $ZZ$ crosstalk reduction, we use a Hamiltonian based on that used in \cite{mundada2019suppression}, which is
\begin{multline}
H/\hbar=\sum_{i=1,2,c}\bigl(q_i\hat{b}_i^\dag\hat{b}_i-\frac{\alpha}{2}\hat{b}_i^\dag\hat{b}_i^\dag\hat{b}_i\hat{b}_i\bigr)\\
    +\sum_{\substack{i,j=1,2,c\\j\neq i}}J_{i,j}\bigl(\hat{b}_i^\dag\hat{b}_j+\hat{b}_i\hat{b}_j^\dag\bigr).
    \label{Houck_Ham}
\end{multline}
Here, $\alpha$ corresponds to the qubit anharmonicity and the $J_{i,j}$ is the qubit-qubit exchange coupling rate between the $i$th and $j$th elements of the circuit. Here, the elements of the circuit are labeled by qubit $1$, qubit $2$, and the SQUID coupler $c$. The $q_i$ are the corresponding energy levels in the context of the Duffing oscillator model. Similarly, $\hat{b}^\dag_i$ and $\hat{b}_i$ are the creation and annihilation operators for the model corresponding to circuit element $i$. 

\begin{figure}[t]
    \centering
    \includegraphics[width=\linewidth]{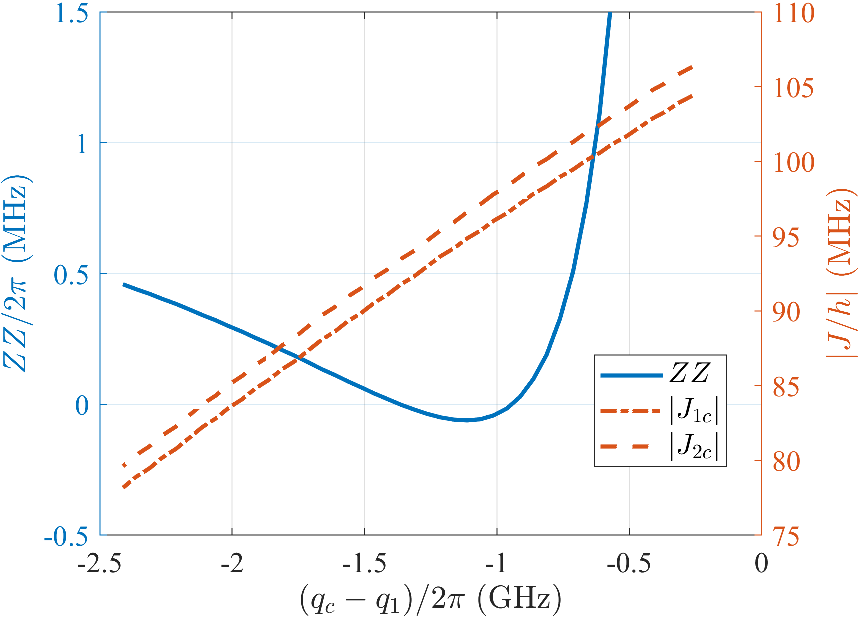}
    \caption{(color online) Section \ref{subsec:Houck} Device: Calculated $ZZ$ rate (solid blue, left axis) from numerically diagonalizing (\ref{Houck_Ham}) while varying the SQUID coupler frequency $q_c$. Calculated values of $J$ rate (dashed and dashed-dot orange lines, right axis) between the qubits and SQUID coupler while varying $q_c$. The computed $ZZ$ rate shows good qualitative agreement with the experimental results of \cite{mundada2019suppression}.}
    \label{Houck_design_Results}
\end{figure}

To evaluate the needed parameters of this Hamiltonian, we use our impedance-based formalism to model the qubits and the SQUID coupler with lumped ports. In the model, we use an effective relative permittivity of $\epsilon_\mathrm{eff}=6.34$ rather than modeling a finite thickness silicon substrate. We use the results of our impedance simulations to extract the total qubit and SQUID coupler capacitances. These are evaluated to be $\mathrm{65.21\,\mathrm{fF}}$ and $\mathrm{65.23\,\mathrm{fF}}$ for the qubits, and $54.95\,\mathrm{fF}$ for the SQUID coupler. These values are used in the numerical routine described in \cite{roth2024maxwell} to compute all needed transition frequencies and anharmonicities for (\ref{Houck_Ham}). We also calculate all the qubit-qubit exchange coupling rates ($J_{12},J_{1c}$ and $J_{2c}$) following the process of the prior sections. To emulate the results of \cite{mundada2019suppression}, we set qubits 1 and 2 at $4.9729\,\mathrm{GHz}$ and $5.1629\,\mathrm{GHz}$, respectively, while varying the frequency of the SQUID coupler from 2.5 to 4.9 GHz. Since the qubits are at fixed frequencies, we get a single value of $J_{12}=-8.82$ MHz, which is primarily dominated due to the coupling effects through the half-wavelength coplanar waveguide resonator. Due to the tuning range of the SQUID coupler, we get a range of values for $J_{1c}$ and $J_{2c}$ which are plotted in Fig. \ref{Houck_design_Results}.

To find the $ZZ$ rate, we use the parameters we have evaluated to numerically build and diagonalize the Hamiltonian from (\ref{Houck_Ham}). This $ZZ$ rate is then computed as $ZZ=E_{11}-E_{10}-E_{01}-E_{00}$, where $E_{mn}$ corresponds to the eigenvalues associated with the dressed eigenstates of the Hamiltonian with $m$ and $n$ excitations in qubits 1 and 2, respectively. The $ZZ$ rate is plotted in Fig. \ref{Houck_design_Results}, which shows good qualitative agreement with the experimental results of \cite{mundada2019suppression}. We note that as we change the frequency of the SQUID coupler $q_c$ the $ZZ$ rate changes and we can find two points $(q_c-q_1)/2\pi \approx-0.96,-1.36\,\,\mathrm{GHz}$ at which the $ZZ$ rate goes to zero. These points are good candidates for the operating location for the multi-qubit device, since they maintain a usable coupling rate for performing multi-qubit operations while effectively eliminating quantum crosstalk effects.

\section{Conclusion}
\label{sec:conclusion}
In this work, we demonstrated how a field-based formalism for cQED systems can be used to derive a simple formula to efficiently evaluate the qubit-qubit exchange coupling rate between transmon qubits. We validated the accuracy of our approach by modeling four different devices and showed good quantitative agreement with 3D numerical diagonalization and to experimental results when full design details were available, as well as good qualitative agreement to experimental results for systems where full design details were not available. Further, we highlighted the robustness of our approach by demonstrating regimes where its accuracy is maintained past that of the related method discussed in \cite{solgun2019simple}.

In the future, our approach can be used for efficiently calculating the qubit-qubit exchange coupling rate for a wide degree of qubit frequencies and detunings, which is essential for designing and better understanding multi-qubit interactions in realistic systems. Our method is also a crucial element to extend efficient Maxwell-Schr\"{o}dinger numerical methods \cite{roth2024maxwell} to accurately model multi-qubit control and readout dynamics in cQED devices. Further, our field-based formalism for cQED systems has elucidated a clear link to the broader field of macroscopic QED. This provides a simple path to leverage the concepts and techniques from macroscopic QED to quickly devise methods to analyze important manifestations of these effects in cQED systems, and as shown here can provide a more transparent and intuitive mathematical derivation than corresponding approaches relying purely on circuit theory.
\begin{acknowledgments}
We acknowledge the help of Soomin Moon in providing their 3D numerical diagonalization method to aid in validating the results of this work. We also appreciate Botao Du and Ruichao (Alex) Ma for providing their four qubit device's design files and consulting on their results. 
\end{acknowledgments}

\appendix
\section{Schrieffer-Wolff Transform}
\label{sec:appendix-SW}
Following the method from \cite{cohen1998atom}, we consider a Hamiltonian system composed of multiple subspaces $\hat{H}=\hat{H}_1\oplus \hat{H}_2 \oplus \ldots \oplus \hat{H}_N$. Our goal is to find an effective Hamiltonian $\hat{H}_{\textrm{eff}}$ that is Hermitian, maintains the original Hamiltonian's eigenenergies, but removes couplings between specific manifolds. That is, $\hat{H}_{\textrm{eff}}$ should satisfy 
\begin{align}
    \label{hod}
    \hat{P}_{\alpha}\hat{H}_{\textrm{eff}}\hat{P}_{\beta}=0 \,\,\,\, \textnormal{for}\;\;\alpha\neq\beta,
\end{align}    
where    
\begin{align}
    \hat{P}_{\alpha}=\sum_{i}\ket{i,\alpha}\bra{i,\alpha}.
\end{align}
These properties can be achieved by using an appropriate unitary transformation 
\begin{align}\label{SW-Unitary}
\hat{H}_{\textrm{eff}}=e^{iS}He^{-iS},
\end{align}
where the generator $S$ is chosen such that 
\begin{align}
    \hat{P}_\alpha \hat{H}_{\textrm{eff}}\hat{P}_\beta=\hat{P}_\alpha e^{i\hat{S}}\hat{H}e^{-i\hat{S}}&\hat{P}_\beta=0\;\;(\alpha\neq\beta),
    \label{SW-cond1}
\end{align}    
\begin{align}
    \hat{P}_\alpha \hat{S}\hat{P}_\alpha&=0.
    \label{SW-cond2}
\end{align}
Here, (\ref{SW-cond1}) sets the off-diagonal elements of $\hat{H}_{\textrm{eff}}$ and (\ref{SW-cond2}) sets the diagonal elements of $\hat{S}$ to zero, allowing the unique determination of $\hat{S}$ and $\hat{H}_{\textrm{eff}}$.

We can then use an auxiliary parameter $\lambda$ for order counting to expand the generator $\hat{S}$ as
\begin{align}\label{SW-expand}
    \hat{S}=\lambda \hat{S}_1 +\lambda^2 \hat{S}_2 + ... + \lambda^n \hat{S}_n.
\end{align}
Next, the original Hamiltonian can be separated into an interacting part $\hat{V}$ and a non-interacting part $\hat{H}_0$ to write $ \hat{H}=\hat{H}_0+\lambda \hat{V}$. We can then use (\ref{SW-expand}) in (\ref{SW-Unitary}) along with the Baker–Campbell–Hausdorff formula to get an expression up to the second order in ${\lambda}$ as
\begin{multline}\label{Eq6}
    \hat{H}_{\textrm{eff}}=\hat{H}_0 + [i\lambda \hat{S}_1,\hat{H}_0]+\lambda \hat{V}\\+[i\lambda^2\hat{S}_2,\hat{H}_0]+[i\lambda \hat{S}_1,\lambda \hat{V}]+\frac{1}{2}[i\lambda \hat{S}_1,[i\lambda{\hat{S}_1},\hat{H}_0]].
\end{multline}
After imposing conditions (\ref{SW-cond1}) and (\ref{SW-cond2}), our expression condenses up to second order terms to be
\begin{align}\label{Eval SW}
    \hat{H}_{\textrm{eff}}=\sum_\alpha\left[\hat{H}_o\hat{P}_\alpha+\hat{P}_\alpha\lambda \hat{V} \hat{P}_\alpha+\frac{1}{2}\hat{P}_\alpha[i\lambda \hat{S}_1,\lambda \hat{V}]\hat{P}_\alpha\right]
\end{align}
and the first-order generator $\hat{S}_1$ can be determined succinctly using 
\begin{align}\label{generator-determiner}
    \bra{i,\alpha}i\hat{S}_1\ket{j,\beta}&=\frac{\bra{i,\alpha}\hat{V}\ket{j,\beta}}{E_{i,\alpha}-E_{j,\beta}}, \textnormal{for }\alpha\neq\beta,\\
    \bra{i,\alpha}i\hat{S}_1\ket{j,\beta}&=0,\nonumber
\end{align}
where $E_{i,\alpha}$ and $E_{j,\beta}$ are the eigenenergies of $\hat{H}_0$ for the states $\ket{i,\alpha}$ and $\ket{j,\beta}$ respectively. We can use (\ref{generator-determiner}) for our Hamiltonian defined in (\ref{original_hamiltonian}) to find the first order generator $\hat{S}_1$ to be
\begin{align}\label{Generator}
    \hat{S}_1=-i\Bigg[\sum_{l=1,2}\sum_{ik}B_{i,k}^{*(l)}\ket{i+1}^{(l)}\bra{i}^{(l)}\hat{a}_k- \textnormal{H.c.}\Bigg]   
\end{align}
where
\begin{align}
    B_{i,k}^{(j)}=\frac{{g}_{i,k}^{(j)}}{q_{i,i+1}^{(j)}-\omega_k}
\end{align}
and $q_{i,i+1}=q_{i+1}-q_{i}$. We can then calculate our effective Hamiltonian using (\ref{Generator}) to get
\begin{widetext}
    \begin{multline}\label{sw-og}         \hat{H}_{\textrm{eff}}=\hbar\sum_jq_j^{(1)}\ket{j}^{(1)}\bra{j}^{(1)}+\hbar\sum_jq_j^{(2)}\ket{j}^{(2)}\bra{j}^{(2)}+\hbar\sum_k \omega_k\hat{a}_k^\dag \hat{a}_k\\
         + \sum_{l=1}^2\left[\hbar\sum_{i,k}\chi_{i,k}^{(l)}\ket{i+1}^{(l)}\bra{i+1}^{(l)}-\hbar\sum_k\chi_{0,k}^{(l)}\hat{a}_k^\dag \hat{a}_k\ket{0}^{(l)}\bra{0}^{(l)}+\hbar\sum_{i=1}^{\infty}(\chi_{i-1,k}^{(l)}-\chi_{i,k}^{(l)})\hat{a}_k^\dag \hat{a}_k\ket{i}^{(l)}\bra{i}^{(l)}\right]\\
         +
          \frac{\hbar}{2}\sum_{i,j,k}\left[\frac{{g}_{i,k}^{*(1)}{g}_{j,k}^{(2)}}{(q_{i,i+1}^{(1)}-\omega_k)}+\frac{{g}_{i,k}^{(1)}{g}_{j,k}^{*(2)}}{(q_{j,j+1}^{(2)}-\omega_k)}\right]\biggl(\ket{i}^{(2)}\bra{i+1}^{(2)}\ket{j+1}^{(1)}\bra{j}^{(1)}+\ket{i+1}^{(2)}\bra{i}^{(2)}\ket{j}^{(1)}\bra{j+1}^{(1)}\biggr),
     \end{multline}
\end{widetext}
 where we have abbreviated the dispersive shifts to the system eigenfrequencies as
 \begin{align}
     \chi_{i,k}^{(l)}=\frac{\lvert {g}_{i,k}^{(l)}\rvert ^2}{q_{i,i+1}^{(l)}-\omega_k}.
 \end{align}

\end{document}